\documentclass[aps,amssymb,amsfonts,twocolumn,showpacs]{revtex4}
\usepackage{graphicx}

\newcommand{\ltsimeq}{\raisebox{-0.6ex}{$\,\stackrel
        {\raisebox{-.2ex}{$\textstyle <$}}{\sim}\,$}}

\begin{document}
\title{Dark soliton decay due to trap anharmonicity in atomic Bose-Einstein condensates}
\author{N.G. Parker$^{1,2}$, N. P. Proukakis $^{1,3}$ and C. S. Adams$^1$}
\affiliation{$^1$ Department of Physics, Durham University, Durham,
DH1 3LE, United Kingdom
\\$^2$ School of Food Science and Nutrition, University of Leeds,
Leeds, LS2 9JT, United Kingdom
\\ $^3$ School of Mathematics and
Statistics, University of Newcastle, Newcastle upon Tyne, NE1 7RU,
United Kingdom}

\begin{abstract}
A number of recent experiments with nearly pure atomic Bose-Einstein
condensates have confirmed the predicted dark soliton oscillations
when under harmonic trapping. However, a dark soliton propagating in
an inhomogeneous condensate has also been predicted to be unstable
to the emission of sound waves. Although harmonic trapping supports
an equilibrium between the co-existing soliton and sound, we show
that the ensuing dynamics are sensitive to trap anharmonicities.
Such anharmonicities can break the soliton-sound equilibrium and
lead to the net decay of the soliton on a considerably shorter
timescale than other dissipation mechanisms. Thus, we propose how
small realistic modifications to existing experimental set-ups could
enable the experimental observation of this decay channel.

\end{abstract}


\maketitle

\section{Introduction}

Dark solitons are one-dimensional phase defects that play an
important role in nonlinear transport.  They have been observed
experimentally in nonlinear optics \cite{krokel}, shallow liquids
\cite{denardo}, magnetic films \cite{chen} and, most recently,
ultra-cold atomic Bose-Einstein Condensates (BECs)
\cite{denschlag,burger,anderson,dutton,weller,becker,stellmer,book}.
In the latter case matter-wave dark  solitons consist of a localised
dip in the atomic density and a phase slip, and are supported by
repulsive atomic interactions.  Dark solitons are prone to several
dissipation mechanisms. Two such mechanisms are thermal dissipation
\cite{fedichev,Muryshev2,jackson_JLTP,jackson2} and the snake
instability whereby the soliton decays into vortex rings
\cite{Muryshev1,carr2000,federnew,brandnew,proukakis_JOptB,Shomroni_note}.
Although numerous initial experiments showed such instabilities to
severely limit the soliton lifetime
\cite{burger,denschlag,dutton,anderson}, they can nonetheless be
heavily suppressed at ultracold temperatures and in highly-elongated
trap geometries. Indeed, such techniques have led to the recent
generation of long-lived dark solitons that made many complete
oscillations in the trapped condensate \cite{weller,becker}. Matter
wave dark solitons are also subject to an instability to
acceleration through an inhomogeneous system
\cite{Busch,Huang,parker_sol_PRL,NOVA,pelinovsky}. We will
demonstrate that this instability is a generic occurrence for dark
solitons in trapped atomic BECs. Furthermore, we will show that,
under appropriately engineered conditions, this can become the
dominant decay mechanism and is within reach of being detected by
current experiments.

Condensate dynamics at zero temperature obey a nonlinear wave
equation known as the Gross-Pitaevskii equation \cite{book}. In
one-dimension and under a homogeneous potential, a dark soliton is
an exact and integrable solution of this equation.  The presence of
an inhomogeneous potential (which is relevant in practically all BEC
experiments) simultaneously accelerates the soliton and breaks this
integrability, rendering it unstable and leading to the emission of
sound waves \cite{Huang}.  Providing the potential is slowly varying
the decay of the soliton energy $E_{\rm s}$ follows a simple
acceleration-squared law
\cite{pelinovsky:pre1996a,kivshar:pr1998,parker_sol_PRL,parker_sol_JPB},
\begin{equation}
\frac{dE_{\rm s}}{dt}=-\kappa \left(\frac{d^2 x_{\rm
s}}{dt^2}\right)^2, \label{eqn:power}
\end{equation}
where $x_{\rm s}$ is the soliton position and $\kappa$ is a
coefficient that typically depends weakly on the local density and
soliton speed. An analogous instability arises for dark solitons in
nonlinear optics, induced by modifications in the optical
nonlinearity, and led to the original derivation of the above power
law \cite{kivshar:pr1998,pelinovsky:pre1996a}.

Importantly, for the common case of harmonic trapping, dark solitons
have been predicted to undergo conservative dynamics in the zero
temperature and one-dimensional limit
\cite{Huang,Busch,frantzeskakis,brazhnyi}, oscillating back and
forth in the trap. This appears to oppose the prediction of
instability under acceleration.  This anomaly can be reconciled by
the fact that in a harmonic trap continuous reabsorption of the
emitted sound stabilises the soliton against decay
\cite{parker_sol_PRL}.  The importance of soliton-sound interactions
was reinforced in \cite{proukakis_pump}, where it was shown that
energy could be pumped into a dark soliton by driving a sound field
in the condensate.  It is interesting to note that a similar
instability arises for a vortex accelerating in a trapped BEC and
that they too appear to be stabilised within harmonic traps
\cite{parker_vor_PRL}.

Harmonic traps appear to be a special case.  Previous results
indicate that deviations from harmonic trapping tend to induce
dissipation of the soliton.  Such instability has been observed, for
example, for a dark soliton in modified harmonic traps
\cite{Busch,parker_sol_PRL}, interacting with localised barriers
\cite{parker_sol_JPB,radouani:pra2003,radouani:pra2004,proukakis_JOptB,frantzeskakis,bilas},
and moving through an optical lattice
\cite{parkerJPBopt,kevrekidis:pra2003} and disordered potential
\cite{bilas2}.  Motivated by this effect we study the soliton
dynamics under several pertinent forms of confining potential.
Crucially, we demonstrate that the presence of anharmonicities can
destroy the soliton-sound equilibrium and lead to net decay of the
soliton. This decay can be rapid and within current experimental
soliton lifetimes, provided the gas is kept at a relatively low
temperature. Importantly, this paper demonstrates that the
experimental observation of this decay mechanism is within reach of
current long-lived soliton experiments
\cite{becker,stellmer,weller}.

We begin in Section II by outlining the theoretical framework and reviewing key
properties of dark solitons and their dynamics in harmonic traps.
Section III identifies various experimentally-relevant geometries for observing/inducing the emission of sound waves, firstly by
introducing a finite cut-off to the harmonic trap,
and then by focusing on soliton dynamics under anharmonic (gaussian; quartic) traps. In Section IV we relate our findings
to the recent long-lived soliton experiments \cite{becker,weller}
and suggest routes to detect this decay mechanism with only minor changes in their
respective experimental set-ups.
Our main conclusions are then summarised in Sec. V.

\section{Background \& Modeling}

\subsection{Gross-Pitaevskii equation}

At ultra-cold temperature and under weak atomic {\it s}-wave
interactions the vast majority of the atoms in an atomic BEC reside
in the same quantum state.  The atomic BEC can then be parameterised
by a macroscopic `wavefunction' $\psi({\bf r},t)$. This can be
expressed as $\psi({\bf r},t)=\sqrt{n({\bf r},t)}\exp[i\theta({\bf
r},t)]$, where $n({\bf r},t)$ is the atomic density distribution and
$\theta({\bf r},t)$ is the condensate phase which can, in turn, be
related to a fluid velocity via $v({\bf r},t)=(\hbar/m)\nabla
\theta({\bf r},t)$. This wavefunction is described by the
Gross-Pitaevskii equation (GPE) \cite{dalfovo},
\begin{eqnarray}
i\hbar \frac{\partial \psi({\bf r},t)}{\partial
t}=\left(-\frac{\hbar^2}{2m}\nabla^2+V_{\rm{ext}}({\bf
r})+g|\psi({\bf r},t)|^2\right)\psi({\bf r},t). \label{eqn:GPE}
\end{eqnarray}
Here $m$ is the atomic mass and $g=4\pi\hbar^2 a_{\rm s}/m$
characterises the nonlinear atomic interactions, where $a_{\rm s}$
is the {\it s}-wave scattering length.  The chemical potential of
the condensate is denoted by $\mu$. The external confining potential
$V_{\rm{ext}}({\bf r})$ is typically harmonic with the form,
\begin{eqnarray}
V_{\rm ext}({\bf
r})=\frac{1}{2}m\left[\omega_r^2\left(x^2+y^2\right)+\omega_z^2z^2
\right], \label{eqn:V}
\end{eqnarray}
where we have assumed cylindrical symmetry, and where $\omega_r$ and
$\omega_z$ are the harmonic trap frequencies in the radial and axial
directions, respectively.  Assuming a highly-elongated condensate
$(\omega_z\ll \omega_r)$ with tight radial confinement
$(\hbar\omega_r>\mu)$, the condensate dynamics become
quasi-one-dimensional and the 1D version of the GPE
\cite{parker_sol_PRL,Huang,Muryshev1,Muryshev2} becomes applicable.
(This also assumes that fluctuations in the condensate phase can be
ignored, which is justified at such low temperatures
\cite{Proukakis_Tutorial}.)

We will simulate the condensate dynamics using the 1D GPE.  This is
performed by numerically propagating the wavefunction $\psi(z,t)$ on
a discrete spatial grid using the Crank-Nicholson technique
\cite{minguzzi}. Although the energy of the soliton technically
extends to infinity, we can numerically evaluate a meaningful local
estimate as outlined in Appendix A.  We present our results in terms
of harmonic oscillator units where time, length and energy are
measured in units of $\omega_z^{-1}$, $l_z=\sqrt{\hbar/m\omega_z}$
and $\hbar \omega_z$, respectively. The profile of the background
condensate on which the soliton is imposed is specified by the ratio
$(\mu/\hbar\omega_z)$. For $\mu \gg \hbar \omega_z$, the system is
in the Thomas-Fermi regime. Here the density profile can be
approximated by $n({\bf r})=[\mu-V({\bf r})]/g$ and the system is
much wider than the soliton.  In the opposite limit of $\mu \ll
\hbar \omega_z$ the system is closer to the non-interacting limit
with size comparable to the soliton size. Throughout this work and
unless otherwise stated we will employ a system defined by $\mu=22.4
\hbar \omega_z$, which is well within the Thomas-Fermi regime.

\subsection{Properties of dark solitons}

In 1D and on a uniform background density $n_0$ a dark soliton with
speed $v$ and position $(z-vt)$ has the analytical form,
\begin{eqnarray}
\psi(z,t)=\sqrt{n_0}e^{-i(\mu/\hbar)t} \left( \beta \tanh \left[ \beta
\frac{\left(z-vt\right)}{\xi}\right]+i\left(\frac{v}{c}\right)\right),
\label{eqn:soliton}
\end{eqnarray}
where $\beta=\sqrt{1-(v/c)^2}$ and the healing length
$\xi=\hbar/\sqrt{m n_0 g}$ characterises the soliton width. For this
homogeneous case the chemical potential is given by $\mu=n_0 g$.
The soliton speed $v$ is related to both the soliton depth $n_d$
(with respect to the background density) and the total phase slip
$S$ across the centre via $v=\sqrt{n_0-n_d}=c\cos(S/2)$, where
$c=\sqrt{n_0 g/m}$ is the Bogoliubov speed of sound. A stationary
soliton has a $\pi$ phase slip and a maximum depth $n_d=n_0$, while
a soliton with speed $c$ is indistinguishable from the background
fluid.

The energy of the dark soliton of Eq.~(\ref{eqn:soliton}), which is
an ideal quantity for parameterising the soliton behavior,  is given
by,
\begin{equation}
E_{\rm s}=\frac{4}{3}\hbar n_0 c\left[1-\left(\frac{v}{c}\right)^2 \right]^{3/2}.
\label{eqn:sol_energy}
\end{equation}
Accordingly, the dissipation of energy from the soliton is
associated with an {\em increase} in its speed. Under an
inhomogeneous potential the soliton behaves, to first order, as an
effective particle of negative mass \cite{DS_particle}.  The soliton
will tend to oscillate when confined within a trap and any
dissipation is associated with an increase in the amplitude of the
soliton oscillations, an effect termed ``anti-damping" \cite{Busch}.
 For the specific case of a harmonic trap a dark soliton is predicted
to oscillate at a frequency $\omega_{\rm s}=\omega_z/\sqrt{2}$
\cite{fedichev,Muryshev1,Busch,Huang,frantzeskakis,pelinovsky,brazhnyi,parker_sol_PRL}.
Note that deviations from this prediction can arise due to the
dimensionality of the system \cite{theocharis2}, the presence of other solitons and
trap anharmonicities, and these effects have been recently observed
experimentally \cite{becker,weller}.

\subsection{Dark solitons in purely harmonic traps}

We first review the dynamics of a dark soliton in a 1D harmonic
trap  $V(z)=m\omega_z^2z^2/2$, which has been the subject of much
theoretical work
\cite{fedichev,Muryshev1,Busch,Huang,frantzeskakis,brazhnyi,parker_sol_PRL}.
The initial state is $\psi(z,0)=\sqrt{n_{\rm TI}(z)} \psi_{\rm
s}(z,0)$, where $n_{\rm TI}(z)$ is the time-independent background
density obtained by imaginary time propagation of the GPE
\cite{imag_time} and the soliton solution $\psi_{\rm s}(z,0)$ is
given by Eq.~(\ref{eqn:soliton}).  The dark soliton solution is
centred at the origin, where the density is locally homogeneous, and
has initial speed $v_0$.
\begin{figure}[t]
\centering
\includegraphics[width=2.5cm,clip=true,angle=-90]{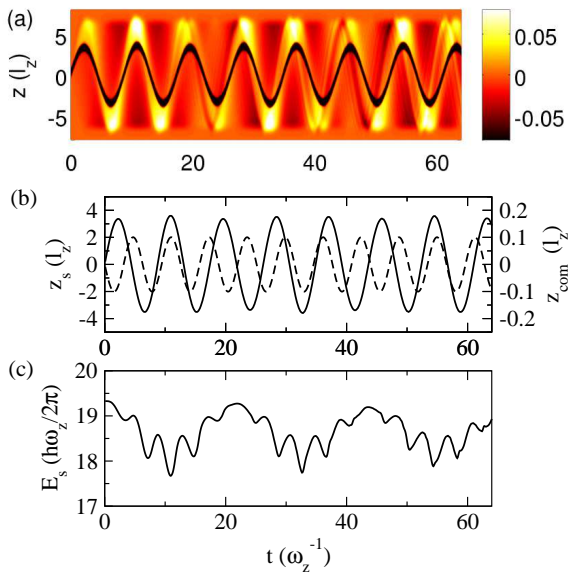}
\hspace{1cm}
\\
\includegraphics[width=7.5cm,clip=true,angle=0]{Fig1b.eps}
\caption{(Color online) Dark soliton dynamics in a
harmonically-trapped condensate for a soliton initially located at the trap centre and with a speed $v_0=0.5c$.   (a) Renormalised density
$|\psi(z,t)|^2-n_{\rm TI}(z)$, in units of the peak density $n_0$.
(b) Soliton position $z_{\rm s}(t)$ (solid line, left axis) and
condensate centre of mass $z_{\rm com}(t)$ (dashed line, right
axis).  The soliton position $z_{\rm s}$ is defined as the point of
the soliton density minimum.  (c) Soliton energy $E_{\rm s}(t)$,
determined numerically as outlined in Appendix A.} \label{fig:sol_harmonic}
\end{figure}

\begin{figure}[t]
\centering
\includegraphics[width=8.5cm,clip=true,angle=0]{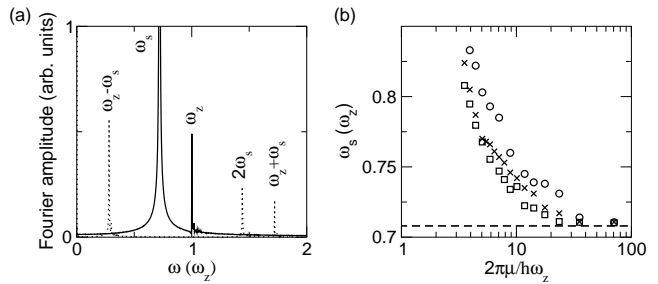}
\caption{(a) Fourier spectrum of $z_{\rm s}(t)$ (solid line) and $E_{\rm s}(t)$ (dotted line) for the case in
Fig.~\ref{fig:sol_harmonic}.
(b) Average soliton frequency $\omega_{\rm s}$ as a function of the system parameter $(\hbar \omega_z/\mu)$ for
$v_0=0.25c$ (squares), $v_0=0.5c$ (crosses) and $v_0=0.75c$ (circles).  The analytic prediction,
$\omega^0_{\rm s}=\omega_z/\sqrt{2}$, is shown (dashed line).} \label{fig:sol_fourier}
\end{figure}

The typical evolution of a dark soliton in a harmonically-confined
BEC is shown in Fig.~\ref{fig:sol_harmonic}(a). The soliton (dark
density dip) oscillates sinusoidally through the BEC.  This is
accompanied by small-scale density excitations of the background
condensate, of amplitude $\sim 0.1n_0$.  We attribute these to the
sound waves emitted from the accelerating soliton.  The sound waves
remain confined within the system and form a dipole (sloshing) mode
of the condensate.
We have tracked the soliton trajectory $z_{\rm s}$ (solid line in
Fig.~\ref{fig:sol_harmonic}(b)) and taken its Fourier spectrum
(solid line in Fig.~\ref{fig:sol_fourier}(a)).  This confirms that
the soliton primarily oscillates at frequency $\omega_{\rm s}
\approx \omega_z/\sqrt{2}$, in agreement with previous predictions
\cite{fedichev,Muryshev1,Busch,Huang,frantzeskakis}.  The spectrum
features a secondary frequency component at $\omega=\omega_z$, which
arises due to the interaction of the soliton with the dipole mode of
the condensate.

During these dynamics the condensate centre-of-mass oscillates due
to the linear momentum introduced by the imposed soliton solution.
The condensate centre-of-mass $z_{\rm com}=\int^{+\infty} _{-\infty}
z |\psi(z)|^2 {\rm d} z/N$ is plotted in
Fig.~\ref{fig:sol_harmonic}(b) (dashed line, right axis).  The
centre-of-mass motion is decoupled from the ``internal'' dynamics of
the condensate and oscillates sinusoidally at the trap frequency
$\omega_z$, as noted in \cite{Busch}.
The soliton energy $E_{\rm s}$  (solid line in
Fig.~\ref{fig:sol_harmonic}(c)), whose definition is outlined in
Appendix A, remains constant on average, but features oscillations.
The corresponding Fourier spectrum (dotted line in
Fig.~\ref{fig:sol_fourier}(a)) has contributions from $2\omega_{\rm
s}$ and beat frequencies between the soliton and dipole mode, e.g.
$(\omega_{z}-\omega_{\rm s})$, $(\omega_{z}+\omega_{\rm s})$, plus
higher order combinations (not visible in Fig.
\ref{fig:sol_fourier}).

The system considered here  is in the
Thomas-Fermi regime $\mu\gg \hbar \omega_z$ (with $\mu=22.4 \hbar \omega_z$).  We have also analysed
the soliton motion over a range of values of $(\mu/\hbar \omega_z)$
with the results shown in Fig.~\ref{fig:sol_fourier}(b).  For
$\mu\gg \hbar \omega_z$ the soliton oscillation frequency
$\omega_{\rm s}$ agrees with the analytic prediction of $\omega_{\rm
s}\approx \omega_z/\sqrt{2}$
\cite{fedichev,Muryshev1,Busch,frantzeskakis}.  Indeed, these
predictions are based on treating the soliton as a classical
particle and so is valid when the soliton is small compared to the
system size.  As the system parameter $(\mu/\hbar\omega_z)$ is
reduced (which reduces the size of the system relative to the
soliton size), $\omega_{\rm s}$ deviates from this prediction and
becomes dependent on the soliton speed.   For $\mu = 5\hbar
\omega_z$, which is when the system size is comparable to the
soliton size, $\omega_{\rm s}$ can be over $10\%$ larger than the
analytic predictions.   Note that in all cases considered, the
soliton dynamics undergo no net decay.  Furthermore, the frequency
component at $\omega_z$ is present throughout the range of systems
tested. Note that shifts in the soliton oscillation frequency can
also arise due to the presence of the transverse direction
\cite{theocharis2,weller} and this is distinct to our present 1D
analysis.

We attribute the density excitations in the system to sound waves
emitted by the unstable accelerating soliton. Although these waves
co-exist with the soliton at long times, it takes a finite time from
$t=0$ for the emitted sound waves to reflect off the edge of the
condensate and reinteract with the soliton.  This will be
approximately half a trap period $t \approx \pi \omega_z^{-1}$.
Close inspection of Fig.~\ref{fig:sol_harmonic}(c) shows that the
soliton energy decays monotonically at these early times.  Indeed,
for this time the decay follows the acceleration-squared law of
Eq.~(\ref{eqn:power}). After this time the soliton energy is found
to oscillate between its initial value and some lower value,
providing a clear indication of the reabsorption of sound by the
soliton.

A recent study by Pelinovsky {\it et al.} \cite{pelinovsky} has
considered the motion of a dark matter-wave soliton in a harmonic
trap by means of an asymptotic multiscale expansion method.  Sound
radiation in the system is predicted to oscillate at the same
frequency as the soliton, thereby causing the periodic
re-interaction  of the sound and soliton, and thus enabling
reabsorption by the soliton.  This prediction is somewhat consistent
with our observations, although we observe the sound waves to have
oscillatory components at the trap frequency as well as the soliton
frequency.

We now discuss various geometries in which the emitted sound waves can be studied.

\section{Probing Anharmonicity-induced Soliton Decay}

\subsection{Dark soliton dynamics in a cut-off harmonic trap}

\begin{figure}[t]
\centering
\includegraphics[width=7.5cm,clip=true,angle=-90]{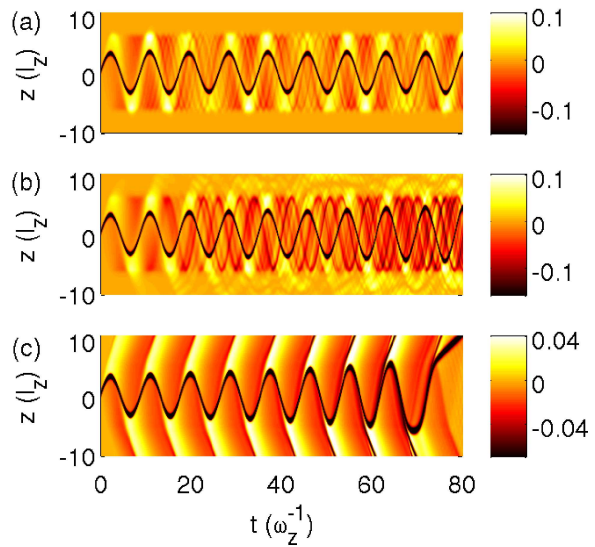}
\caption{ (Color online) Space-time plots of renormalised density
($|\psi(z,t)|^2-n_{\rm TI}(z)$) showing the dynamics of a $v_0=0.5c$
dark soliton in cut-off harmonic trap with (a) $V_0/\mu=1.2$, (b)
$1.05$ and (c) $0.8$.  The dark soliton appears as a black line,
while sound waves appear as light/dark areas.}
\label{fig:cutoff_carpets}
\end{figure}

\begin{figure}[t]
\centering
\includegraphics[width=8.5cm,clip=true]{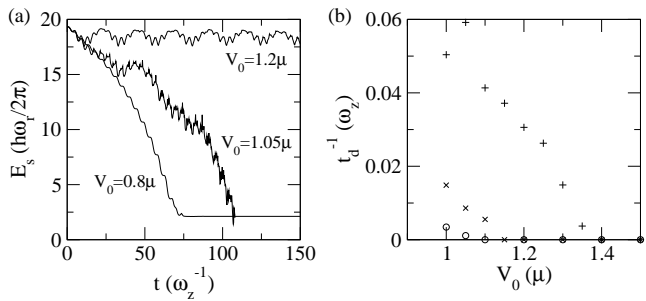}
\caption{(a) Evolution of the soliton energy $E_{\rm s}$ in a
cut-off harmonic trap for $V_0/\mu=0.8$, $1.05$ and $1.2$ for
initial speed $v_0=0.5c$. (b) Decay rate $t_{\rm d}^{-1}$ as a
function of trap cut-off $V_0$ for $v_0=0.25c$ (circles), $v_0=0.5c$
(crosses) and $v_0=0.75c$ (pluses).} \label{fig:cutoff_decay}
\end{figure}

Although a purely harmonic trap generates no net decay of the dark soliton,
we can nonetheless probe the existence of sound emission even within harmonic traps by introducing a
finite cut-off to the trap, as discussed in \cite{parker_sol_PRL}.
We thus consider a cut-off harmonic trap of
the form,
\begin{equation}
V_{\rm ext}(z)=\left\{
\begin{array}{c}
m\omega_{z}^2z^2/2 \\
 V_0
\end{array}
\right.
{\rm for}
\begin{array}{cr}
& |z|< \sqrt{2V_0/m\omega_z^2} \\
& |z|> \sqrt{2V_0/m\omega_z^2}
\end{array}
\end{equation}
where $V_0$ specifies the magnitude of the cut-off.  Although this
potential is somewhat unphysical, it can be approximated by
employing a tight dimple trap embedded within a weaker ambient trap
(i.e.\ an effective double-trap geometry) \cite{parker_sol_PRL}.
Since sound waves have an energy of approximately the chemical
potential $\mu$, the cut-off depth enables control between a regime
where the sound waves remain confined to the inner trap
($V_0\gg\mu$) and a regime where they can escape the inner trap
($V_0<\mu$).  In Fig.~ \ref{fig:cutoff_carpets} we show the
evolution of the system for various values of $V_0$ which illustrate
these regimes.    For $V_0=1.2\mu$
(Fig.~\ref{fig:cutoff_carpets}(a)) the sound waves are confined to
the harmonic region and the evolution is essentially identical to
that under pure harmonic trapping (Fig.~\ref{fig:sol_harmonic}(a)).
For this case the soliton energy, presented in
Fig.~\ref{fig:cutoff_decay}(a), shows no net decay. For $V_0=0.8\mu$
(Fig.~\ref{fig:cutoff_carpets}(c)) all of the emitted sound waves
escape the harmonic region and propagate outwards.
The soliton energy
(Fig.~\ref{fig:cutoff_decay}(a)) decays monotonically and follows
the acceleration-squared law of Eq.~(\ref{eqn:power})
\cite{parker_sol_PRL}.  Note that at late times the soliton has
anti-damped sufficiently that it escapes to the outer homogeneous
region where it propagates with constant speed. In between these
regimes, e.g. for $V_0=1.05\mu$ (Fig.~\ref{fig:cutoff_carpets}(b)),
only a fraction of the sound waves escape to infinity, and we
observe a slower decay of the soliton.  Note that the power law of
Eq.~ (\ref{eqn:power}) no longer applies since considerable sound
reabsorption occurs.  The emitted sound remains mostly confined to
the central trap and the amplification of this sound field as the
soliton decays is clearly evident.

For the cut-off harmonic trap, decay of the soliton occurs because
sound waves escape the harmonic region and carry away their energy.
For $V_0<\mu$ this can always happen.  However, it is important to
understand how decay changes as $V_0$ becomes greater than $\mu$, as this will
prove crucial in our subsequent discussion on soliton decay in experimentally relevant gaussian traps. We
define a decay time $t_{\rm d}$ as the time for the soliton to
become indistinguishable from the background condensate.  In
Fig.~\ref{fig:cutoff_decay}(b) we show how the decay rate $t_{\rm
d}^{-1}$ varies as $V_0$ becomes greater than $\mu$.  The most
important thing to note is that the decay rate drops very quickly
with $V_0$.  For the soliton speed $v_0=0.5c$, which we have
considered in Fig. \ref{fig:cutoff_carpets}, decay ceases for
$V_0\geq 1.15\mu$.  This corresponds to the point at which sound
waves no longer escape the harmonic region.  This threshold is
larger for faster solitons.  For example, for $v_0=0.75c$, decay is
ultimately prevented for $V_0\geq 1.4\mu$.  This suggests that the
sound waves emitted by faster solitons are of higher energy. However
there always exists a threshold cut-off depth beyond which the tendency to decay is suppressed.

Having reviewed and extended previous results of dark soliton dynamics in harmonic traps we now progress to consider
an important additional decay mechanism for dark solitons, namely
their propagation in condensates under anharmonic confinement.

\subsection{Dark soliton dynamics in a gaussian trap}

For simplicity, and in order to make contact with realistic experiments,
we begin by considering a gaussian trap; such traps are
commonly used experimentally and are formed in the cross-section of
an off-resonant laser beam.  Notably, the recent dark soliton
experiment of Becker {\it et al.} \cite{becker} was performed in a
trap which was gaussian in the longitudinal direction. We will focus
on this particular experiment further in Section V.  However, in
this section we will present the general dynamics in a gaussian
trap.

We consider here a 1D gaussian trap of the form,
\begin{eqnarray}
V_{\rm ext}(z)&=&V_0\left[1-\exp \left(\frac{-m\omega_z^2
z^2}{2V_0}\right)\right]
\\
&=&\frac{m}{2}\omega_z^2 z^2 -\frac{V_0}{2}\left( \frac{m\omega_z^2
z^2}{2V_0}\right)^2+\frac{V_0}{3}\left(\frac{m\omega_z^2
z^2}{2V_0}\right)^3+. . .\nonumber
\label{eqn:1D_gaussian_trap}
\end{eqnarray}
The leading term in the Taylor expansion is harmonic, with trap
frequency $\omega_z$, while the higher order terms are anharmonic.
In a similar manner to the cut-off trap, the parameter $V_0$
controls whether sound waves can escape the system.  However, we
will generally consider the regime $V_0>\mu$, where the condensate
(including soliton and sound) is wholly confined to the gaussian
trap. Importantly, the gaussian depth $V_0$ also controls the degree
of anharmonicity experienced by the condensate, and can be varied
experimentally by varying the laser intensity. In the limit $V_0 \gg
\mu$, the confinement is effectively harmonic, while for
$V_0\sim\mu$ the condensate will experience strong anharmonic terms.

\begin{figure}[t]
\centering
\includegraphics[width=7.5cm,clip=true,angle=-90]{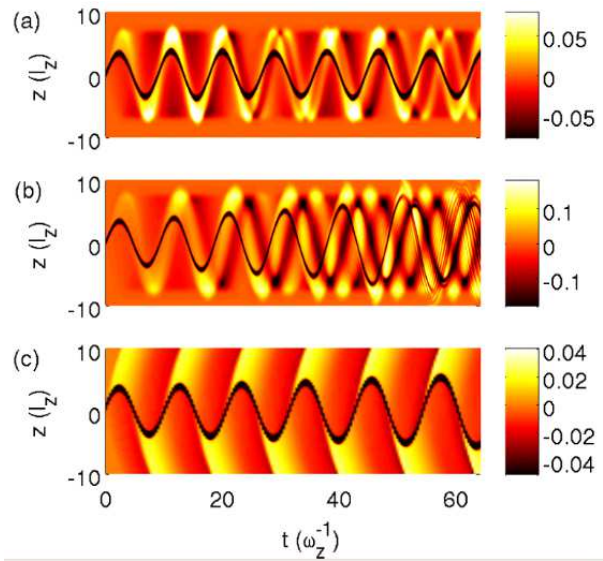}
\caption{Space-time plots of renormalised density
($|\psi(z,t)|^2-n_{\rm TI}(z)$) showing the dynamics of a $v_0=0.5c$
dark soliton in a (a) deep gaussian trap with $V_0=5\mu$, (b)
shallow gaussian trap with $V_0=2\mu$, and (c) infinite system with
$V_0=0.8\mu$.} \label{fig:1D_carpets}
\end{figure}

\begin{figure}[t]
\centering
\includegraphics[width=8cm,clip=true]{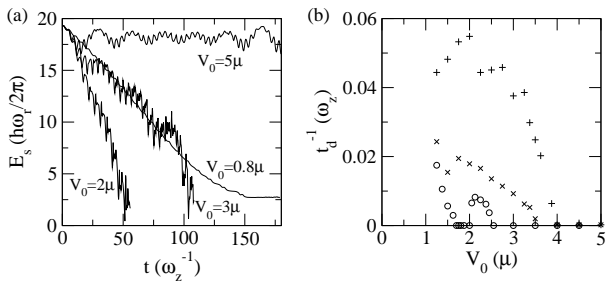}
\caption{(a) Evolution of the soliton energy $E_{\rm s}$ in a
gaussian trap for various depths $V_0$ for a soliton with initial
speed $v_0=0.5c$. (b) Decay rate $t_{\rm d}^{-1}$ as a function of
$V_0$ for $v_0=0.25c$ (circles), $v_0=0.5c$ (crosses) and
$v_0=0.75c$ (pluses).} \label{fig:soliton_decay}
\end{figure}

We first consider how the gaussian depth affects the soliton
dynamics, with Fig.~\ref{fig:1D_carpets} showing the density
evolution for three values of $V_0$. For $V_0=5\mu$
[Fig.~\ref{fig:1D_carpets}(a)], the trapping is effectively harmonic
and the dynamics are similar to those in a harmonic trap, with the
soliton energy, shown in Fig.~\ref{fig:soliton_decay}(a), undergoing
no net decay.  As for the purely harmonic trap this occurs due to
the complete reabsorption of the emitted sound by the soliton.

In contrast, for $V_0=2\mu$ (Fig.~\ref{fig:1D_carpets}(b)),  the
soliton amplitude increases and its energy decreases
(Fig.~\ref{fig:soliton_decay}(a)). The soliton decay is accompanied
by the growth of sound waves in the system.  After only 5
oscillations ($t\approx 50 \omega_z^{-1}$) the soliton is so fast
and shallow that it is indistinguishable from the sound field.
It is important to note here that the
trap depth is sufficiently large that no sound waves escape the
system and so the decay mechanism is quite distinct from that of the
shallow cut-off harmonic trap.  While the sound waves remain
confined to the trap and co-exist with the soliton, they do not
stabilise the decay. We believe that the anharmonic nature of the
trap introduces a dephasing between the sound and soliton that
prevents the sound from being completely reabsorbed.
For completeness we also present the dynamics for $V_0=0.8\mu$
(Fig.~\ref{fig:1D_carpets}(c)) where sound waves can escape the
trap.  In this case we observe the soliton undergoing decay with no
sound reabsorption, similar to the corresponding case for the
cut-off harmonic trap (Fig.~\ref{fig:cutoff_carpets}(c)).

In Fig.~\ref{fig:soliton_decay}(b) we plot the decay rate $t_{\rm
d}^{-1}$ of the soliton versus the gaussian depth $V_0$ for various
soliton speeds. Consider the case of $v_0=0.5c$ (crosses).  Here
decay occurs up to $V_0=3.5\mu$. Note that according to the results
of Fig.~\ref{fig:cutoff_decay}(b) we expect that no sound can escape
for $V_0>1.5\mu$ and so it is clear that the decay occurs even in
the presence of the emitted sound.  We see that the decay rate
decreases with $V_0$.  This is consistent with the decay being due
to anharmonicities: as $V_0$ increases the trap becomes increasingly
harmonic (from the perspective of the condensate).  Furthermore we
see that faster solitons have larger decay rates and undergo decay
up to larger values of $V_0$. We attribute this to the fact that
faster solitons probe closer to the edge of the condensate where the
trap anharmonicities have the largest effect.

It is important to note that the soliton decay in the gaussian trap
can be considerable.  For example, for the case presented in
Fig.~\ref{fig:1D_carpets}(b), the soliton has practically
disappeared by $t=50 \omega_z^{-1}$, which corresponds to
approximately $5$ soliton oscillations in the trap. Assuming a
typical trap frequency of $\omega_z=2\pi \times 100~$Hz, this
timescale corresponds to around $80~$ms, which is well within the
thermodynamic lifetime of dark solitons at ultralow temperatures
\cite{jackson2,weller,becker}. Furthermore, the fastest decay
presented in Fig. \ref{fig:soliton_decay}(b) occurs twice as fast as
this example, corresponding to only two or three soliton
oscillations.

\subsection{Dark Soliton Dynamics in a Quartic Trap}

\begin{figure}[t]
\centering
\includegraphics[width=6cm,clip=true]{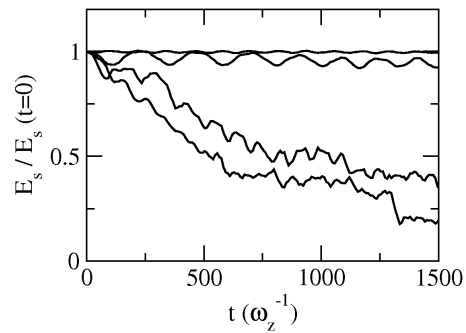}
\caption{(a) Evolution of the soliton energy $E_{\rm s}$ in a
quartic trap for soliton speeds of (from top to bottom) $v_0/c=0.5$,
$0.6$, $0.7$ and $0.8$. The trap has the form of
Eq.~(\ref{eqn:quartic_trap}) with $V_0=10\mu$. } \label{fig:quartic}
\end{figure}

To give further evidence of the
role of trap anharmonicities in dark soliton decay we briefly
consider the soliton dynamics within a quartic trap. Quartic
potentials can be obtained, for instance, by modifying the gaussian
potential from a laser beam, as performed experimentally to create
harmonic-plus-quartic traps \cite{bretin}. As such, we assume our
quartic trap to have the form of the quartic term of the gaussian
potential (Eq.~\ref{eqn:1D_gaussian_trap}), i.e.,
\begin{equation}
 V_{\rm ext}(z)=\frac{1}{2 V_0} \left(\frac{m\omega_z^2 z^2}{2} \right)^2.
 \label{eqn:quartic_trap}
\end{equation}
Following previous sections, we define our system via
$\mu=22.4\hbar\omega_z$, and assume here $V_0=10\mu$. In
Fig.~\ref{fig:quartic} we present the evolution of the soliton
energy for various soliton speeds.  We observe that the soliton
stability is strongly dependent on speed. Slow solitons
$v_0\leq0.5c$ undergo little or no detectable decay, while faster
solitons, e.g. $v_0=0.8c$, do undergo decay. Note that the decay is
significantly slower than in the gaussian trap.

We have also considered further traps formed by a combination of
harmonic and quartic terms, and observe regimes of soliton decay in
all cases.  For a harmonic-plus-quartic trap, the trap has similar
properties to the purely quartic trap and we obtain similar
qualitative results. For a harmonic-minus-quartic trap, the trap is
similar to a gaussian trap, particularly in the sense that it only
has finite depth, and we find similar results to the gaussian trap.

We now discuss how our findings relate to recent experiments conducted at very low temperatures.

\section{Application to recent experiments}

In the early dark soliton experiments
\cite{denschlag,burger,anderson,dutton} the solitons were short
lived due to thermal dissipation and the snake instability. This
limited the observation times to a fraction of the soliton
oscillation period. Recent experiments \cite{becker,weller,stellmer}
have heavily suppressed these dissipation mechanisms by operating at
extremely cold temperatures and by employing quasi-1D geometries. As
a result they were able to observe solitons over much longer
timescales in which the soliton undergoes multiple oscillations in
the trap. Indeed, in the experiment of Becker {\it et al.}
\cite{becker} dark solitons were observed for over $2.8~$s,
corresponding to approximately $10$ soliton oscillations in the
trap.  The considerable soliton lifetimes in these recent
experiments, as well as the suppression of thermal dissipation and
the snake instability, offer realistic opportunities to
experimentally detect the acceleration-induced decay of the soliton
discussed here.

\subsection{Experiment of Becker et al.}

We firstly consider the experiment of Becker {\it et al.} \cite{becker}.  Dark
solitons were generated in an elongated $^{87}$Rb condensate via a
phase imprinting technique and observed to oscillate in the trap
with peak velocity $v_0=0.56c$. The trap, formed by the optical
dipole force of an applied laser beam, featured a gaussian profile
in the axial direction with waist $W_z=125~\mu$m and dominant
harmonic frequency $\omega_z=2\pi \times 5.9$ Hz.  For the
experimentally-quoted peak density of $n_0=5.8 \times 10^{13}~{\rm
cm}^{-3}$ we estimate the chemical potential of the condensate to be
$\mu/k_B \sim n_0 g /k_B \sim 23~$nk.  The amplitude of the gaussian
potential is $V_0=m\omega_z^2 W_z^2/4=58~$nK.  The important
depth-to-chemical potential ratio is then $V_0/\mu \sim 2.5$. Based
on these parameters our simulation gives a soliton oscillation
frequency of $\omega_{\rm s}=2\pi \times 3.94~$Hz and an amplitude
of $Z_0 \approx 34\mu$m, which are both in good agreement with the
values quoted in Ref. \cite{becker}.

\begin{figure}[t]
\centering
\includegraphics[width=8.5cm,clip=true]{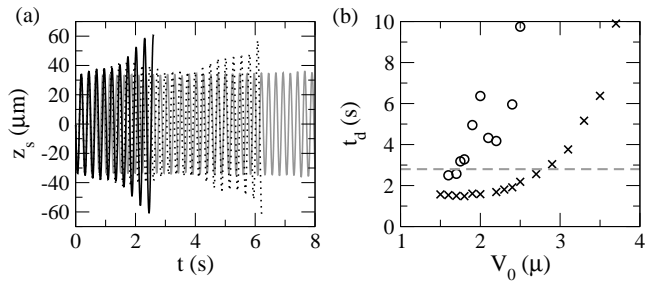}
\caption{Soliton decay in the experimental system of
Ref.~\cite{becker}. (a) The evolution of the soliton position for
$V_0=1.7\mu$ (solid line), $V_0=2\mu$ (dotted line) and $V_0=4\mu$
(grey line).  Following the experimental observations we employ an
initial soliton speed of $v_0=0.56c$. (b) Decay time in the
experimental system as a function of the gaussian trap depth $V_0$.
Initial soliton speeds of $v_0=0.56c$ (circles) and $v_0=0.75c$
(crosses) are presented.  The dashed line indicates the maximum
soliton lifetime reported in the experiment.} \label{fig:hamburg}
\end{figure}

It is pertinent to ask whether acceleration-induced decay of the
soliton occurs in this system, and whether it may be detectable. In
Fig.~\ref{fig:hamburg}(a) we plot the soliton trajectory in the
experimental system for several values of the trap depth $V_0$. Note
that the trap depth $V_0$ is proportional to, and can be varied by
modifying, the intensity of the laser beam that forms the trap.  For
$V_0=4\mu$ (grey line) the trap is essentially harmonic: the soliton
undergoes no discernable decay and oscillates with constant
amplitude. For $V_0=2\mu$ (dotted line), which is just smaller than
the number used in the experiments, the soliton experiences weak
trap anharmonicities, causing it to decay slowly (as evident from
the increase in the soliton amplitude) and effectively disappear by
$t\sim 6~$s. This decay timescale is longer than the reported
experimental lifetime ($2.8~$s), and so evidence of such decay would
not be easily visible in the current experiment. Under a slight
reduction in the trap depth to $V_0/\mu=1.7$ (solid line) the
soliton undergoes much quicker decay, disappearing into the
background condensate after only $2.5$ s.   Importantly, this is now
within the experimental soliton lifetime.

The effect of trap depth is shown in more detail in
Fig.~\ref{fig:hamburg}(b) which plots the decay time as a function
of trap depth $V_0$ for two different initial soliton speeds. The
maximum reported experimental soliton lifetime ($2.8~$s) is shown
(dashed line) for reference.  As earlier results indicated, the
decay time generally increases with $V_0$ as the trap becomes
increasingly harmonic.  For the experimental trap depth $V_0\sim
2.5\mu$ and soliton speed $v_0=0.56c$ the soliton decays over a
timescale $t_{\rm d} \approx 6~$s, which is longer than the
experimental lifetime. How could one modify the experiment to make
the decay occur within the experimental timescale? Firstly, this can
be achieved (assuming an initial soliton speed of $v_0=0.56c$) by
reducing the trap depth to the range $V_0\ltsimeq 1.8\mu$. Secondly,
the decay can be made more rapid by beginning with a faster soliton.
For example, for $v_0=0.75c$ the soliton will decay within the
experimental lifetime provided the trap depth does not exceed $V_0
\sim 2.8\mu$.

These results suggest that with only slight modifications this decay
channel will dominate the experiment and cause the soliton to decay
within the experimental timescale. The soliton dynamics are also
very sensitive to the trap depth $V_0$, a feature which also lends
itself to experimental detection.  The detectable signature for the
decay would be provided by the considerable anti-damping of the
soliton that accompanies the decay, which could be probed either in
expansion images \cite{burger,denschlag,weller,becker,stellmer} or
even in situ \cite{shomroni}.

\subsection{Experiment of Weller et al.}

We will also briefly discuss the recent soliton experiment by Weller
{\it et al.} \cite{weller}.  Two separated condensates were
initially formed in a double trap potential.  The double trap
potential is formed by the combination of an elongated harmonic trap
and one period of a one-dimensional optical lattice potential.  Upon
removal of the optical lattice the condensate underwent nonlinear
interference which resulted in the production of dark solitons.
These solitons were then observed to make up to $7$ oscillations in
the system. Although the solitons are typically produced in pairs we
expect our predictions to hold for multiple solitons.  The final
trap potential is harmonic and so we do not expect any net decay of
the solitons due to acceleration-induced sound emission. However, if
the optical lattice potential were instead maintained at some low
amplitude following the creation of the dark solitons, then we
anticipate that soliton decay would occur.

We previously considered the dynamics of a dark soliton in a
harmonic trap featuring an optical lattice potential
\cite{parkerJPBopt}. The optical lattice potential was assumed to be
weak (amplitude less than the chemical potential) such that the
soliton could traverse the lattice site/s and thereby undergo
acceleration-induced sound emission.  Provided the lattice spacing
$\lambda$ was in the range $\xi<\lambda<2L_z$, where $\xi$ is the
healing length and $L_z$ is the condensate length, the soliton
consistently underwent decay. (For $\lambda<\xi$ the density
perturbations due to the optical lattice become smeared out while
for $\lambda>2L_z$ the trap reduces to being effectively harmonic.)
This is consistent with the results presented in this paper: the
optical lattice introduces anharmonicities that prevent complete
sound reabsorption. The ensuing decay could be rapid, with complete
decay occurring as quickly as within one trap period in certain
cases. In the experiment of Weller {\it et al.} \cite{weller} the
lattice spacing is approximately $L_z$, and so we expect that the
presence of the lattice can indeed induce soliton decay. This
suggestion is similar to an earlier prediction
\cite{proukakis_JOptB} where we showed that a single gaussian
barrier within a harmonic trap could lead to soliton decay.
Moreover, decay in the experiment of \cite{weller} could be further
enhanced by decreasing the periodicity of the proposed optical
lattice; following previous results \cite{parkerJPBopt} we would
expect the quickest decay when $\lambda \sim 5 \xi$.

\section{Discussion}

We have examined the zero-temperature, mean-field dynamics of dark
solitons in atomic Bose-Einstein condensates under harmonic and
anharmonic trapping.  Such inhomogeneous potentials generally
accelerate the soliton and induce an instability whereby the soliton
radiates sound waves.  In the absence of sound reabsorption the
power radiated by the soliton is proportional to its acceleration
squared.  However, due to the finite-size of BECs, the sound waves
remain in the system and generally re-interact with the soliton. The
ensuing dynamics between the soliton and sound waves are crucially
dependent on the form of the trapping potential.

Under harmonic trapping no net decay of the soliton is observed,
despite the local instability of the soliton. The isochronicity of
the harmonic trap leads to the creation of a dynamical equilibrium between
the soliton and the emitted sound waves which prevents net soliton
decay.  We note that the inclusion of an additional harmonic
frequency to the system, e.g., by considering an asymmetric harmonic
trap (different trap frequencies for $z>0$ and $z<0$), maintains the
stability of the soliton.  Decay of the soliton in the harmonic trap
can be induced, but involves removal of some of the emitted sound,
e.g. by introducing a finite cut-off to the harmonic trap.

In a gaussian trap decay can occur without the removal of the
emitted sound even at zero temperature.  This is possible because
the soliton-sound equilibrium is broken by the anharmonicity of the
trap. In accord with this the decay is strongly dependent on the
anharmonicity experienced by the soliton. The soliton decay is
accompanied by the growth of short wavelengths in the sound field.
Quartic trapping also leads to soliton decay. Furthermore, several
existing studies have added some form of anharmonic perturbation to
the harmonic trap and observed soliton decay, e.g. under the
addition of an optical lattice
\cite{parkerJPBopt,kevrekidis:pra2003}, a localised barrier
\cite{proukakis_JOptB,frantzeskakis} and a disordered potential
\cite{bilas2}, in concurrence with our findings.

Soliton decay can be rapid.  For example, in the gaussian trap we
have observed a soliton (with initial speed $v_0=0.5c$, where $c$ is
the speed of sound) disappear into the sound field after only five
oscillations in the trap.  Given that solitons have recently been
stabilised for over $7$ trap oscillations \cite{weller,becker} it is
clear that acceleration-induced soliton decay can dominate over the
more well-known dissipation mechanisms (thermal dissipation and the
snake instability).  For the experiment of Becker et al.
\cite{becker}, which was conducted in a gaussian trap, our results
suggest that the soliton should completely decay within the
experimental lifetime provided a slightly shallower gaussian trap or
a faster soliton is employed. Note that the trap depth is readily
varied through the intensity of the laser beam that forms the trap,
while the soliton speed can be varied by adjusting the degree of
phase imprinting. Furthermore, for the recent experiment of Weller
{\it et al.} \cite{weller} we suggest that if a weak optical lattice
is maintained during soliton propagation it may lead to significant
soliton decay. Soliton decay can be detected by the increase in the
soliton oscillation amplitude measured either in successive
absorption images or in situ, with the accompanying growth of the
sound field providing further indication of this effect.

It is interesting to note that the instability to acceleration that
underpins this decay channel has also been predicted for vortical
excitations in atomic BECs, including single vortices
\cite{lundh,vinen,parker_vor_PRL}, corotating vortex pairs
\cite{pismen} and two-dimensional vortex-antivortex pairs
\cite{barenghi}.  Hybrid dark soliton/vortex rings excitations that
have recently been observed \cite{shomroni,ginsberg} may also be
expected to undergo an instability to acceleration.  As such the
decay channel discussed here may extend well beyond dark solitons.

\begin{acknowledgments}
We acknowledge the UK EPSRC (NGP/NPP/CSA) and the Canadian
Government (NGP) for support.  We thank A. M. Martin, D. H. J.
O'Dell, D. Pelinovsky and K. Sengstock for discussions.
\end{acknowledgments}

\appendix
\section{Evaluation of the dark soliton energy}
\begin{figure}[b]
\includegraphics[width=3.8cm,clip=true,angle=-90]{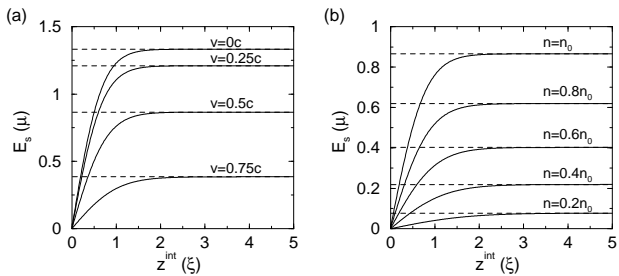}
\caption{Integrated local soliton energy $E_{\rm s}$ as a function
of integration width $z^{\rm int}$. (a) Various soliton speeds are
considered in a system with peak density $n=n_0$. (b) Various values
of background density $n$ are considered for fixed soliton speed
$v=0.5c$.  In each case the analytic soliton energy is indicated by
a dashed line.  This is performed in an infinite $V(z)=0$ system and
our units of energy and distance are the chemical potential $\mu=n_0
g$ and healing length $\xi=\hbar\sqrt{m n_0 g}$ of the $n=n_0$
system.} \label{soliton_energy}
\end{figure}

Although the soliton energy technically extends up to the boundary
of the system it is useful to have a {\em local} measurement of the
soliton energy.  We perform this by numerically evaluating the
energy in a local region about the soliton centre.  In trapped
systems this will typically contain contributions from other
excitations, e.g. sound waves. However, this contribution is
typically very small.  Note that from an experimental point of view
it is precisely the behaviour of this soliton region, including
local excitations, which is probed experimentally.

The energy density of the condensate is given by the energy functional,
\begin{eqnarray}
\varepsilon[\psi]=\frac{\hbar^2}{2m}\left|\nabla\psi\right|^2+V_{\rm
ext}\left|\psi\right|^2+\frac{g}{2}\left|\psi\right|^4.\label{eqn:func}
\end{eqnarray}
We define our soliton energy $E_{\rm s}$ to be,
\begin{equation}
E_{\rm s}=\int_{z_{\rm s}-z_{\rm int}}^{z_{\rm s}+z_{\rm int}}
\varepsilon[\psi] {\rm d} z- \int_{z_{\rm s}-z_{\rm int}}^{z_{\rm
s}+z_{\rm int}} \varepsilon[\sqrt{n_{\rm TI}}] {\rm d}z.
\label{eqn:soliton_energy}
\end{equation}
Here $z_{\rm int}$ defines the extent of the integration region.  Note that we subtract off the corresponding energy of the condensate background density (second term).  Figures \ref{soliton_energy}(a) and (b) present the integrated
soliton energy as a function of $z_{\rm int}$ for various soliton speeds and background densities, respectively.  For simplicity we consider an infinite system with $V(z)=0$.  For large enough $z_{\rm int}$ we observe convergence
to the analytic soliton energy of Eq.~\ref{eqn:sol_energy} (dashed lines). Larger $z_{\rm int}$ is required for convergence for faster solitons
and/or lower densities, due to the increased soliton width.
However, for the speeds and densities considered here, the integrated energy is virtually indistinguishable from the
analytic prediction for $z^{\rm int}= 4\xi$, which is employed in this work.


\begin{thebibliography}{99}
\bibitem{imag_time} Using an appropriate initial guess, propagation of the GPE in imaginary
time will converge to the ground state of the system, provided it
exists \cite{minguzzi}.
\end{thebibliography}
\end{document}